\documentclass[journal=nalefd,manuscript=letter]{achemso}

\usepackage{color}
\usepackage{amsmath}

\author{Giuseppe Toscano}
\affiliation[Center for Nanoscience and Nanotechnology, School of Physics and Technology, and Institute for Advanced Studies, Wuhan
University, Wuhan 430072, China]{Center for Nanoscience and Nanotechnology, School of Physics and Technology, and Institute for Advanced Studies,
Wuhan University, Wuhan 430072, China}
\alsoaffiliation[Institut f\"{u}r Theoretische Festk\"{o}rperphysik, Karlsruhe Institute of Technology (KIT), D-76131 Karlsruhe,
Germany]{Institut f\"{u}r Theoretische Festk\"{o}rperphysik, Karlsruhe Institute of Technology (KIT), D-76131 Karlsruhe, Germany}
\email{giuseppe.toscano@kit.edu}

\author{Jakob Straubel}
\affiliation[Institut f\"{u}r Theoretische Festk\"{o}rperphysik, Karlsruhe Institute of Technology (KIT), D-76131 Karlsruhe,
Germany]{Institut f\"{u}r Theoretische Festk\"{o}rperphysik, Karlsruhe Institute of Technology (KIT), D-76131 Karlsruhe, Germany}

\author{Alexander Kwiatkowski}
\affiliation[Institut f\"{u}r Theoretische Festk\"{o}rperphysik, Karlsruhe Institute of Technology (KIT), D-76131 Karlsruhe,
Germany]{Institut f\"{u}r Theoretische Festk\"{o}rperphysik, Karlsruhe Institute of Technology (KIT), D-76131 Karlsruhe, Germany}

\author{Carsten Rockstuhl}
\affiliation[Institut f\"{u}r Theoretische Festk\"{o}rperphysik, Karlsruhe Institute of Technology (KIT), D-76131 Karlsruhe,
Germany]{Institut f\"{u}r Theoretische Festk\"{o}rperphysik, Karlsruhe Institute of Technology (KIT), D-76131 Karlsruhe, Germany}
\alsoaffiliation[Institute of Nanotechnology, Karlsruhe Institute of Technology (KIT), Hermann-von-Helmholtz-Platz 1,
D-76021 Eggenstein-Leopoldshafen, Germany]{Institute of Nanotechnology, Karlsruhe Institute of Technology (KIT),
Hermann-von-Helmholtz-Platz 1, D-76021 Eggenstein-Leopoldshafen, Germany}

\author{Ferdinand Evers}
\affiliation[Institut I - Theoretische Physik, Universit\"{a}t Regensburg, Universit\"{a}tsstra{\ss}e 31, D-93053 Regensburg, Germany]{Institut I
- Theoretische Physik, Universit\"{a}t Regensburg, Universit\"{a}tsstra{\ss}e 31, D-93053 Regensburg, Germany}
\alsoaffiliation[Institute of Nanotechnology, Karlsruhe Institute of Technology (KIT), Hermann-von-Helmholtz-Platz 1,
D-76021 Eggenstein-Leopoldshafen, Germany]{Institute of Nanotechnology, Karlsruhe Institute of Technology (KIT),
Hermann-von-Helmholtz-Platz 1, D-76021 Eggenstein-Leopoldshafen, Germany}

\author{Hongxing Xu}
\affiliation[Beijing National Laboratory for Condensed Matter Physics and Institute of Physics, Chinese Academy of Sciences,
Beijing 100190, China]{Beijing National Laboratory for Condensed Matter Physics and Institute of Physics, Chinese Academy of
Sciences, Beijing 100190, China}
\alsoaffiliation[Center for Nanoscience and Nanotechnology, School of Physics and Technology, and Institute for Advanced Studies, Wuhan
University, Wuhan 430072, China]{Center for Nanoscience and Nanotechnology, School of Physics and Technology, and Institute for Advanced Studies,
Wuhan University, Wuhan 430072, China}

\author{N. Asger Mortensen}
\affiliation[DTU Fotonik, Department of Photonics Engineering, Technical University of Denmark, DK-2800 Kgs. Lyngby,
Denmark]{DTU Fotonik, Department of Photonics Engineering, Technical University of Denmark, DK-2800 Kgs. Lyngby, Denmark}
\alsoaffiliation[Center for Nanostructured Graphene (CNG) , Technical University of Denmark, DK-2800 Kgs. Lyngby, Denmark]{Center
for Nanostructured Graphene (CNG) , Technical University of Denmark, DK-2800 Kgs. Lyngby, Denmark}

\author{Martijn Wubs}
\affiliation[DTU Fotonik, Department of Photonics Engineering, Technical University of Denmark, DK-2800 Kgs. Lyngby,
Denmark]{DTU Fotonik, Department of Photonics Engineering, Technical University of Denmark, DK-2800 Kgs. Lyngby, Denmark}
\alsoaffiliation[Center for Nanostructured Graphene (CNG) , Technical University of Denmark, DK-2800 Kgs. Lyngby, Denmark]{Center
for Nanostructured Graphene (CNG) , Technical University of Denmark, DK-2800 Kgs. Lyngby, Denmark}

\title[An \textsf{achemso} demo]
{Resonance shifts and spill-out effects in self-consistent hydrodynamic nanoplasmonics}
\abbreviations{IR,NMR,UV}

\begin{document}

%\begin{tocentry}

%Some journals require a graphical entry for the Table of Contents.
%This should be laid out ``print ready'' so that the sizing of the

%Inside the \texttt{tocentry} environment, the font used is Helvetica
%8\,pt, as required by \emph{Journal of the American Chemical

%The surrounding frame is 9\,cm by 3.5\,cm, which is the maximum
%permitted for  \emph{Journal of the American Chemical Society}
%graphical table of content entries. The box will not resize if the
%content is too big: instead it will overflow the edge of the box.

%This box and the associated title will always be printed on a
%separate page at the end of the document.

%\end{tocentry}

\begin{abstract}
The standard hydrodynamic Drude model with hard-wall boundary conditions can give accurate quantitative predictions
for the optical response of noble-metal nanoparticles. However, it is less accurate for other metallic nanosystems, where surface
effects due to electron density spill-out in free space cannot be neglected.
Here we address the fundamental question  whether the description of surface
effects in plasmonics necessarily requires a fully quantum-mechanical approach,
such as time-dependent density-functional theory (TD-DFT), that goes beyond an
effective Drude-type model. We present a more general formulation of the
hydrodynamic model for the inhomogeneous electron gas, which additionally
includes gradients of the electron density in the energy functional. In doing so, we arrive at a
Self-Consistent Hydrodynamic Model (SC-HDM), where spill-out emerges naturally. 
We find a redshift for the optical response of
Na nanowires, and a blueshift for Ag nanowires, which are both in
quantitative agreement with experiments and more advanced quantum methods. The
SC-HDM  gives accurate results with modest computational effort, and can be
applied to arbitrary nanoplasmonic systems of much larger sizes than accessible
with TD-DFT methods. Moreover, while the latter typically neglect retardation
effects due to time-varying magnetic fields, our SC-HDM takes retardation
fully into account.
\end{abstract}

\section{Introduction}
Many recent experiments on dimers and other structures in nanoplasmonics have revealed phenomena beyond classical electrodynamics, but their nature is not always clear and often debated. To explain these new phenomena,  a unified theory would be highly desirable that includes nonlocal response, electronic spill-out, as well as retardation. And ideally such a theory would lead to an efficient concomitant computational method. To his end, one could try to  generalize standard density-functional theory to include retardation, or to generalize standard hydrodynamic theory so as to include free-electron spill-out at metal-dielectric interfaces. In this Letter we present the latter, both the generalized hydrodynamic theory and its efficient numerical implementation.

The hydrodynamic Drude model of the electron gas is one of the most commonly used methods for the computational study of plasmonic
nanostructures beyond classical electrodynamics~\cite{Raza:2011}. This model has been used in
recent years to determine the optical response of coupled
nanoparticles~\cite{Mcmahon:2009,Christin:2011,Toscano:2012a,Mittra:2012,
Antonio:2012b,Antonio:2012,Ciraci:2013,Luo:2013,Wiener:2013}, the
limitations on field enhancement in subwavelength
regions~\cite{Toscano:2012b,Ciraci:2012}, the nanofocusing performances of
plasmon tips~\cite{Wiener:2012}, and the fundamental limitations of Purcell
factors in plasmonic waveguides~\cite{Toscano:2013}. The most important feature
of this model is its ability to describe the plasmonic response of noble metals
nanoparticles at size regimes at which the classical Drude metal is no longer
valid, with a high accuracy and a low computational effort. This is obtained by
including the Thomas--Fermi pressure in the equation of
motion of the electron gas~\cite{Raza:2011} that takes into account the fermion
statistics of the electrons. The hydrodynamic Drude model model provides accurate
predictions of the spectral positions of the surface plasmon resonances (SPR) for
noble-metal dimers, and for single particles it mimics the SPR blueshift that
grows as the particle size decreases.

In line with the local-response Drude model, the standard implementation of the
hydrodynamic Drude model considers the so-called \emph{hard-wall boundary
condition}, that implies
that the electrons are strictly confined in the metallic structure, with a
uniform equilibrium density and without spill-out in free
space~\cite{Raza:2011}. This approximation is more accurate for noble
metals, that show a high work function~\cite{Lang:1971}, than for alkali metals~\cite{Brack:1993,Weick:2006,Stella:2013,Teperik:2013}, where the electron density spill-out in free-space characterizes the plasmonic response.
This is a well-known limitation of the hydrodynamic Drude
model~\cite{Apell:1984}, and many attempts have been made to overcome it, thus
far with limited success.

In a pioneering attempt,~\citeauthor{Schwartz:1982}
\cite{Schwartz:1982} proposed to relax the hard-wall constraint by introducing
equilibrium electron-density distributions obtained with different methods in
the hydrodynamic equations. They found that continuous surface densities lead
to the proliferation of spurious surface multipole modes, qualitatively
disagreeing with experimental results, which lead them to conclude that the HDM
approach to surface effects is subject to a considerable uncertainty. This was
disputed by~\citeauthor{Zaremba:1994}\cite{Zaremba:1994}, who pointed out that
the equilibrium density of the electrons in a metal cannot be chosen
arbitrarily, but it must be calculated self-consistently within the hydrodynamic
theory itself.

In this Letter we introduce a novel self-consistent hydrodynamic theory for the inhomogeneous electron gas.  Hard-wall or other additional boundary conditions are no longer needed and electronic spill-out emerges naturally. We apply this method to the study of the optical response of alkali (Na) and noble-metal (Ag) nanowires, because sodium nanowires were recently used as an example to illustrate the lack of accuracy of standard hydrodynamic theory~\cite{Stella:2013,Teperik:2013}. By contrast, without any fitting and without heavy computations we obtain with our self-consistent hydrodynamic method accurate agreement both with experiments and with more advanced theories.
This illustrates the promising use of our self-consistent hydrodynamic method in nanoplasmonics,
where efficient methods that take into account retardation, nonlocal response, as well as spill-out are highly demanded but scarce.

\section{Hydrodynamic equations}
We first briefly recall how in general hydrodynamic equations are derived  in
terms of functional derivatives of the internal energy of an electron plasma,
before specifying our choice for the internal energy. The Bloch hydrodynamic
Hamiltonian for the inhomogeneous electron gas~\cite{Bloch:1933} reads:
\begin{align}\label{equ:Hamiltonian}
H[n({\bf r},t),{\bf p}({\bf r},t)]= G[n({\bf r},t)] + \int \frac{\left( {\bf
p}({\bf r},t)-e {\bf A}({\bf r},t) \right)^2}{2m} n({\bf r},t) d{\bf r} + e
\int \phi({\bf r},t) n({\bf r},t) d{\bf r} + \nonumber & \\ + e \int V_{\rm back}({\bf r}) n({\bf r},t)d{\bf
r},
\end{align}
where $n({\bf r},t)$ is the electron density and ${\bf p}({\bf r},t)=m {\bf
v}({\bf r},t)+ e {\bf A}({\bf r},t)$ its conjugate momentum, with $e$ being
the electron charge and $m$ the electron mass. The electrons are
coupled to the electromagnetic field associated with the retarded
potentials $\phi({\bf r},t)$ and ${\bf A}({\bf r},t)$, i.e.
${\bf E}({\bf r},t) = - \nabla \phi({\bf r},t) - \frac{\partial {\bf A}({\bf
r},t)}{\partial t}$ and ${\bf B}({\bf r},t) = \nabla \times {\bf A}({\bf
r},t)$. The electrostatic potential $V_{\rm back}({\bf r})$ is a confining
background potential, that is associated with the electrostatic field generated
by the positive ions in a metal, i.e. $\nabla^2 V_{\rm back}({\bf r}) = -
\rho^+({\bf r})/\varepsilon_{0}$, where $\rho^+({\bf r})$ is  the positive
charge density of the  metal ions. The term $G[n({\bf r},t)]$ is the internal
energy of the electron gas, which is assumed to be a functional of the electron
density. Its functional form is not known \textit{a priori} and we will
specify and motivate our particular choice later. The equations of motion can be
obtained from eq~\ref{equ:Hamiltonian} by means of the methods of the
Hamiltonian formulation of
fluid dynamics~\cite{Morrison:1998,Greene:1980,Morrison:2005} (see Supplemental
Material).

The force balance on a fluid element is stated by the Euler equation [and in
the following we will for simplicity suppress the $({\bf r},t)$  dependence]
\begin{equation}\label{equ:Euler}
m n \Big ( \frac{\partial {\bf v}}{\partial t} + {\bf v} \cdot \nabla {\bf v} \Big )=- n
\nabla
\frac{\delta G}{\delta n} + n e ({\bf E} + {\bf v} \times {\bf B}),
\end{equation}
where the density $n$ satisfies the charge continuity equation
\begin{equation}\label{equ:continuity}
\frac{\partial n}{\partial t}=-\nabla \cdot ( n{\bf v}).
\end{equation}

We focus on the linear response of the electron gas described by eq~\ref{equ:Hamiltonian}, so we
solve eqs \ref{equ:Euler} and \ref{equ:continuity} for small perturbations of the electron density $n_1({\bf r},t)$
around the equilibrium density $n_{\rm 0}({\bf r})$, i.e. $n({\bf
r},t)=n_0({\bf r})+n_1({\bf r},t)$ \cite{Eguiluz:1976}. It can be shown (see
Supplemental Material) that $n_{\rm 0}({\bf r})$ satisfies
\begin{equation}\label{equ:n_0}
\Big ( \frac{\delta G}{\delta n} \Big )_0 + e ( \phi_0 + V_{\rm back}) = \mu,
\end{equation}
where $\phi_0$ is the potential associated with the electrostatic field ${\bf
E}_0$ generated by the equilibrium charge density
$\rho_0=e n_0$. Thus $\phi_0$ and $\rho_0$ are related by Poisson's equation
$\nabla^2 \phi_0({\bf r}) = - \rho_0({\bf
r})/\varepsilon_{0}$. The quantity $\big (\frac{\delta G}{\delta n} \big )_0$ is
the the functional derivative %$\frac{\delta G}{\delta n}$
evaluated at the equilibrium density $n_0$,
%, i.e. $\frac{\delta G}{\delta n}\big|_{n_0}$.
and $\mu$ represents the (constant) chemical
potential of the electron gas.  Equation~\ref{equ:n_0} is quite general and also appears in standard density
functional theory~\cite{Parr_Yang_book,Hohenberg:1964}.

It is useful to write the linearized versions of both the force-balance
eq~\ref{equ:Euler} and the charge continuity eq~\ref{equ:continuity} in terms
of the electric-charge density perturbation $\rho_1=e n_1$, and the first-order
electric current-density vector ${\bf J}_1=e n_0 {\bf v}_1 \equiv \rho_0 {\bf
v}_1$. The linearized Euler equation for the current-density vector ${\bf
J}_1$ reads
\begin{equation}\label{equ:linear_Euler}
\frac{\partial {\bf J}_1}{\partial t}=- \frac{\rho_0}{m} \nabla \Big ( \frac{\delta G}{\delta
n} \Big )_1 + \omega_{\rm p}^2 \varepsilon_0 {\bf E}_1,
\end{equation}
while the linearized continuity equation becomes
\begin{equation}\label{equ:linear_continuity}
\nabla \cdot {\bf J}_1=-\frac{\partial \rho_1}{\partial t}.
\end{equation}
The quantity $\big ( \frac{\delta G}{\delta n} \big )_1$
in eq~\ref{equ:linear_Euler} is given by $\frac{\delta G}{\delta
n}\big|_{n_0+n_1}-\frac{\delta
G}{\delta n}\big|_{n_0}$, and
$\omega_{\rm p}=
[e^2 n_0/(m \varepsilon_{0})]^{1/2}$ is the common plasma frequency of the electron gas.

The vector fields ${\bf E}_1$ and ${\bf J}_1$ satisfy Maxwell's wave equation
\begin{equation}\label{equ:Maxwell}
\nabla \times \nabla \times {\bf E}_1 + \frac{1}{c^2} \frac{\partial^2{\bf E}_1 }{\partial
t^2}=-\mu_0 \frac{\partial {\bf J}_1}{\partial t}.
\end{equation}
The linear system given by eqs~\ref{equ:linear_Euler}, \ref{equ:linear_continuity}, and \ref{equ:Maxwell} is closed,
and can be
solved once the equilibrium density $\rho_0({\bf r})$ has been calculated by
means of eq~\ref{equ:n_0} (see Supplemental Material). It is important to notice
that retardation effects due to time-varying magnetic fields are included in
the system above, so that both longitudinal and transverse solutions are
possible. This is an important difference with state-of-the-art TD-DFT methods,
that only solve for longitudinal currents and electric
fields~\cite{Ullrich:1991}.

The linearization procedure of the hydrodynamic equations that we have outlined
so far is logically consistent, since no external assumptions on the quantities
involved are required. In particular, rather than a postulated function,
%typically postulated constant inside the metal and vanishing outside,
the static electron density $n_{0}({\bf r})$ is a solution of eq.~\ref{equ:n_0}.  We call this method ``Self-Consistent Hydrodynamic Model'' (SC-HDM), in order to distinguish it from other implementations of the
hydrodynamic model that employ an arbitrarily chosen equilibrium charge
density. And for our choice of the energy functionals as detailed below, we will be able to describe electronic spill-out, i.e. a non-vanishing $n({\bf r})$  also outside the metal.

\subsection{Hydrodynamic Drude model}
The hydrodynamic Drude model commonly
used in nanoplasmonics can be presented in terms of the general hydrodynamic
theory developed above. We recall here the equation
of motion for this model, that in time domain reads\cite{Raza:2011}
\begin{align}\label{equ:equ_motion_HDM}
\frac{\partial {\bf J}_1}{\partial t}=- \beta^2 \nabla \rho_1 + \omega_{\rm
p}^2 \varepsilon_0 {\bf E}_1,
\end{align}
where $\beta$ is called \emph{hydrodynamic parameter}, defined as
$\beta=\sqrt{3/5}v_{\rm F}$, and $v_{\rm F}$ is the Fermi velocity, $v_{\rm
F}=\frac{\hbar}{m} (3 \pi^2)^{1/3}n_0^{1/3}$. In the usual
implementations of the hydrodynamic Drude model, a uniform equilibrium
electron density $n_0(\mathbf{r})=n_0$ is assumed, and
the electron density spill-out in free space is neglected. This leads to the
\emph{hard-wall boundary condition}, stating that the current density vector
${\bf J}_1$ has a vanishing normal component at the surface of the metallic
nanostructures. For this reason, in the next sections we will refer to this
model as ``Hard-Wall Hydrodynamic Model'' (HW-HDM).

It can be readily shown that eq \ref{equ:equ_motion_HDM} can be derived from eq
\ref{equ:linear_Euler}, if we consider $n_0(\mathbf{r})=n_0$ inside the metal, and approximate
the internal energy $G[n]$ by means of the Thomas--Fermi kinetic energy
functional multiplied by a constant parameter $\alpha=9/5$ that account for the high frequency corrections \cite{Halevi:1995} (see below the discussion about the choice of the energy
functionals). Moreover, it is possible to prove that a uniform
equilibrium electron density is a solution of eq \ref{equ:n_0}, thus
the hydrodynamic Drude model is a special case of the SC-HDM
(see Supplemental Material).

Finally, we would like to mention that if the hydrodynamic parameter
$\beta$ vanishes, eq \ref{equ:equ_motion_HDM} reduces to the Drude equation of
motion for the free-electron gas. We call this well-known model ``Local Response
Approximation'' (LRA), since it neglects nonlocal electromagnetic effects
associated with the derivative of the charge density $\rho_1$ in the equation of
motion.

\subsection{Including density gradients in the energy functional}
After obtaining the closed system of linearized hydrodynamic equations, we now
 specify and motivate the form of the energy functional that features in
them. The internal energy functional $G[n({\bf r},t)]$ in
eq~\ref{equ:Hamiltonian} is given, in general, by the sum of a kinetic
energy functional $T[n({\bf r},t)]$ and an exchange-correlation (XC)
energy functional $F_{\rm xc}[n({\bf r},t)]$~\cite{Hohenberg:1964}:
\begin{equation}\nonumber
G[n({\bf r},t)]=T[n({\bf r},t)]+F_{\rm xc}[n({\bf r},t)].
\end{equation}
In TD-DFT, the kinetic-energy term $T[n({\bf r},t)]$ is
calculated by means of the Kohn--Sham equations, a task that is computationally
demanding for particles consisting of much more than $10^3$ electrons, in other
words for most experiments in nanoplasmonics. The main goal of our SC-HD model
is therefore to provide an alternative computational tool for
particles and structures of larger sizes and lower symmetry than TD-DFT
currently can handle. We do this by means of an orbital-free
approach~\cite{Vignale:1997, Banerjee:2000, Vignale_book, Neuhauser:2011}. Below we discuss the kinetic- and
XC-energy functionals separately.

We approximate the kinetic-energy functional  by  the sum of the
Thomas--Fermi functional and the von Weizs\"{a}cker functional~\cite{Yang:1986},
\begin{equation}\label{equ:T}
T_{\rm TFW}[n]=T_{\rm TF}[n] + T_{\rm W}[n] = \frac{3}{10} \frac{\hbar^2}{m}
(3 \pi^2)^{2/3} \int
n^{5/3}({\bf r},t)\,d{\bf r} + \frac{1}{72} \frac{\hbar^2}{m} \int \frac{|\nabla n({\bf r},t)|^2 }{n({\bf r},t)} \,d{\bf
r}.
\end{equation}
Here the last term is the von Weizs\"{a}cker functional, which is also called
the ``second-order gradient correction'' to the Thomas--Fermi kinetic
functional~\cite{Parr_Yang_book}. Such a gradient correction can become important
in space regions where $n_0({\bf r})$ shows strong variations, for example in
the vicinity of metal surfaces as we shall see below, and especially when describing molecules\cite{Parr_Yang_book}. Our choice is motivated by the fact that the $T_{\rm TFW}$ functional is known to provide reliable descriptions of surface effects in metals, and it has been extensively employed in solid-state physics to calculate the work functions
and surface potentials of various metals~\cite{Smith:1969,Chizmeshya:1988,Costantino:1987,Tarazona:1989,
Snider:1983}. Although adding the functional $T_{\rm TFW}$ will significantly
improve the accuracy of hydrodynamic calculations, it will not
reconstruct the wave nature of the electrons, inherent e.g. in the Friedel
oscillations into the metal bulk, which require the full Kohn--Sham
calculations with XC functionals~\cite{Sorbello:1983}.

In nanoplasmonics only the Thomas--Fermi functional was used until
now~\cite{Mcmahon:2009,Christin:2011,Toscano:2012a,Mittra:2012,Antonio:2012b,
Antonio:2012,Toscano:2012b,Ciraci:2012,Ciraci:2013,Luo:2013,Wiener:2013,
Wiener:2012,Toscano:2013}, while to our knowledge the second-order gradient
correction has thus far been neglected. Below we will give an example
where calculations of optical properties based on only the Thomas--Fermi
functional disagree with experiment, while agreement is found when including the
von Weizs\"{a}cker functional, in combination with the XC functionals as
discussed below.

An exact expression for the XC functional is not known in
general~\cite{Martin_book}, and it is not easy to calculate it numerically
either, so an approximation of the $F_{\rm xc}[n({\bf r},t)]$ is commonly
employed, for example for doing TD-DFT calculations. The most familiar one is
the local-density approximation (LDA), where the name implies that no density
gradients are considered in the construction of the functional. Just like for the kinetic functional,
gradient corrections to the XC functional may nonetheless become significant for inhomogeneous systems. This
calls for  nonlocal density approximations (NLDA) beyond LDA, in order to
correct for long-range effects arising for strong density variations.

On a par with our treatment of the kinetic functional, we will also include
density gradients in the XC-functional that enters our hydrodynamic equations.
In particular, we include the Gunnarson and Lundqvist (GL) LDA
XC functional~\cite{Gunnarson:1976}, that has been used both
for Na~\cite{Ekardt:1984,Teperik:2013} and Ag~\cite{Pustovit:2006}.
Additionally, we include a non-local correction (NLDA) for long-range effects, namely the van Leeuwen and Baerends
potential (LB94)~\cite{Leeuwen:1994},  which has already been
used for Ag and Au plasmonic nanostructures~\cite{Guidez:2013,Piccini:2013}. Incidentally, when leaving out the NLDA exchange term to the functional, we found non-physical solutions with density tails propagating in the free-space region.

Hereby we have specified and motivated the total energy functional that we will employ in our hydrodynamic theory.  The inclusion of the density-gradient terms will make the crucial difference with state-of-the-art hydrodynamic theory used in plasmonics. We will soon show predictions that agree well both with experiments and with more advanced theory. Yet we want to point up that we do not claim to have identified the unique and only functionals. For example, more refined gradient corrections to the kinetic energy functionals have been proposed in the literature\cite{Skriver:1998,DellaSala:2011,DellaSala:2014}, and they might be needed in order to approximate the DFT results for specific cases.

%For more details and a brief comparison of our functional with the one of Vignale, Ulrich and Conti %(refs.~\citenum{Vignale:1997,Vignale_book}) we refer to the Supplementary Material.

%%%%%%%%%%%%%%%%%%%%%%%%%%%%%%%%%%%%%%%%%%%%%%%%%%%%%%%%%%%%%%%%%%%%%%%%%%%%%%%%%%%%%%
\subsection{Implementation of the SC-HD model}
We implemented the SC-HD model with the finite-element method in the
commercially available software COMSOL Multiphysics 4.2a.
Equation~\ref{equ:n_0} is nonlinear in the electron density, and it can be
easily implemented when the ionic background of the metallic system is
specified. We employ the jellium model for the ions, a standard approximation
that is also successfully adopted in state-of-the-art TD-DFT
calculations~\cite{Teperik:2013,Stella:2013,Kulkarni:2013}.

The linear differential system given by eqs~\ref{equ:linear_Euler} and
\ref{equ:Maxwell} was transformed to the frequency domain. We then solved for
the induced charge density $\rho_1$ instead of the induced current-density
vector ${\mathbf J}_1$. This is numerically advantageous, since it allows to
solve for only one variable
($\rho_1$) instead of the three components of ${\mathbf J}_1$. The
full implementation procedure is described in detail in the Supplemental
Material.

\section{Sodium Nanowires - Optical response}
\begin{figure}[ht]
\centering \includegraphics{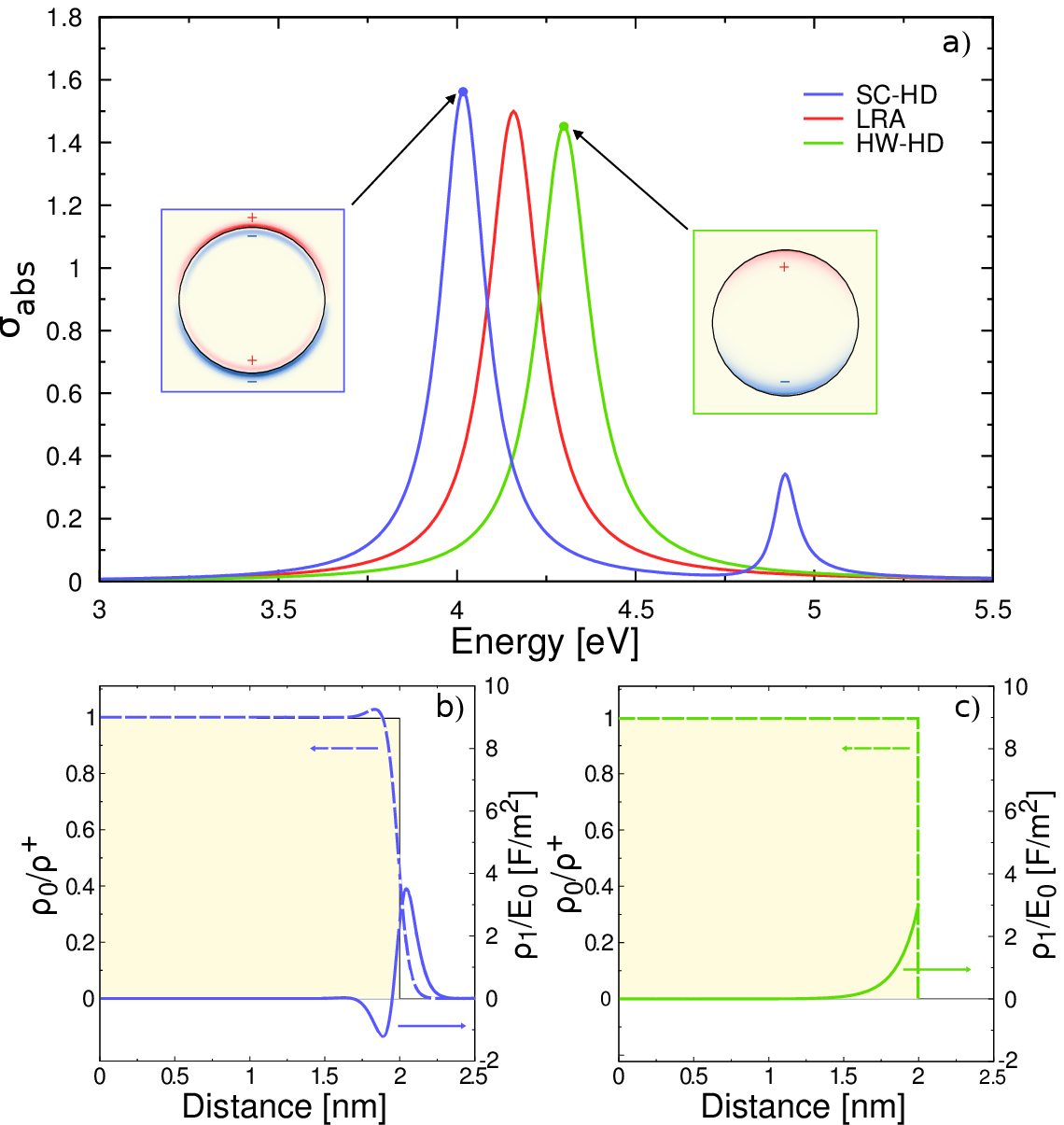}
\caption{Panel a) Absorption cross section $\sigma_{\rm abs}$ versus photon
energy for a Na cylindrical nanowire of radius
$R=2$\,nm in the LRA (red line), the HW-HD model (green line) and the SC-HD model (blue line).
The insets show the charge density distributions $\rho_{\rm 1}(\bf r)$ at their
respective SPR frequencies for the SC-HD model (blue box) and the HD
model (green box). Panel b) SC-HDM model: equilibrium electron charge density
$\rho_{\rm 0}$ normalized to the ion charge density $\rho^{\rm +}$ (dashed
line), and induced charge density
$\rho_{\rm 1}$ normalized to the perturbing field amplitude $E_{\rm 0}$ for the
SPR, as a function of the distance from the axis of the nanowire. Panel c) As
panel b), now for the HW-HD model.}
\label{fig:sodium}
\end{figure}

We apply our formalism to derive the optical response of a Na cylindrical
nanowire of radius $R= 2 \, {\rm nm}$. In the first step of our analysis, we
calculate the equilibrium electron density $n_0$ of the
Na nanowire by means of eq~\ref{equ:n_0}. The positive background density of
the jellium is given by the ion density of Na, i.e. $n^+=2.5173 \times 10^{28}
\, {\rm m^{-3}}$. The result is shown in fig~\ref{fig:sodium}b (dashed line).
Besides the charge spill-out into free space, $n_{\rm 0}$ also exhibits an
oscillation within the metal. This is somewhat reminiscent of Friedel
oscillations, whereas true Friedel oscillations at wavelength $\lambda_{\rm F}/2$
with the correct envelope require solving the full
Kohn--Sham equation with suitable XC functionals.

To determine the optical response of the system, we irradiate the nanowire
with an in-plane polarized plane wave of amplitude $E_{\rm 0}=1 \, {\rm V/m}$,
as in ref~\citenum{Toscano:2012a}. All the perturbed quantities are calculated
by means of the equation of motion eq~\ref{equ:linear_Euler}, where a damping
factor $\gamma$ is  introduced to take into account the electron-phonon
interaction, dissipation due to the electron-hole continuum, impurity
scattering, etc. We choose the value $\hbar\gamma=0.17 \, {\rm eV}$ from
ref~\citenum{Palik:1991}. Interband effects are not included, since for Na their
effects in the visible range are negligible~\cite{Marder:2010}.

The optical quantity that we consider is the absorption cross section per unit
length $\sigma_{\rm abs}$, defined as $\sigma_{\rm abs}=\frac{P_{\rm abs}}{2 R
I_{\rm 0}}$, where $I_{\rm0}= \varepsilon_{0} c E_{\rm 0}^2/2$ is the power
density of the perturbing field. We compare the prediction by the SC-HD model
with
both the HW-HDM and with the usual local-response approximation (LRA). The
hydrodynamic parameter $\beta$ for the HW-HDM is
$\beta=\sqrt{3/5}v_{\rm F}$, where the Fermi velocity for Na is $v_{\rm F}=0.82
\times 10^6 \, {\rm m/s}$.

The results are shown in Figure~\ref{fig:sodium}a. The main SPR within
the LRA occurs at $\hbar\omega_{\rm res}=4.158 \, \rm eV$. This
resonance is {\em blueshifted} by $0.139\,\rm eV$ to become $\hbar\omega_{\rm
res}=4.297\,\rm eV$ in the HW-HDM. In stark contrast, for the SC-HD
model the resonance appears at $\hbar\omega_{\rm res}=4.017 \, {\rm eV}$, in
other words {\em redshifted} by $0.141\,\rm eV$ with respect to the LRA.

Apart from their different peak positions, the three SPRs in the
absorption cross section in Figure~\ref{fig:sodium}a also exhibit different
peak heights in the three models. An intuitive explanation for this behavior
could be provided on the basis of the different geometrical
cross sections in the three models: in the SC-HDM the electrons spill out from
the surface of the nanowire, so the structure appears bigger than in the LRA.
In
contrast, the free electrons  are pushed inwards into the metal in the
HW-HD model, so this structure looks smaller than in the LRA.

The redshift of the surface-plasmon resonance in Na is linked to the
electron spill-out in free space, or more precisely, to the fact that the
centroid of the induced charge density $\rho_1$ associated with the SPR is
placed outside the jellium edge~\cite{Weick:2006,Teperik:2013,Liebsch:1993}.
Our self-consistent calculations correctly reproduce this feature of $\rho_{\rm
1}$ at the SPR peak, as it is shown in Figure~\ref{fig:sodium}b (filled line).
This
plot highlights a peak of $\rho_{\rm
1}$ in the free-space region and a dip in the internal
region
where the equilibrium density takes its maximal value. We
normalized the induced charge density $\rho_{\rm 1}$
with respect to $E_{\rm 0}$, since in the considered linear-response regime the
induced charge density is linearly proportional to the amplitude of the applied
field.

To further confirm the validity of our method, we would like to stress that the
spectral position $\hbar\omega_{\rm res}=4.017 \, {\rm eV}$ of the SPR that we
obtained within our semiclassical hydrodynamic theory, is in good agreement with the values obtained with TD-DFT methodologies for similar systems~\cite{Stella:2013,Teperik:2013}. A perfect agreement is not achievable because of the different approximations that are intrinsic to each specific method.

\begin{figure}[ht]
\centering \includegraphics[width=0.8\textwidth]{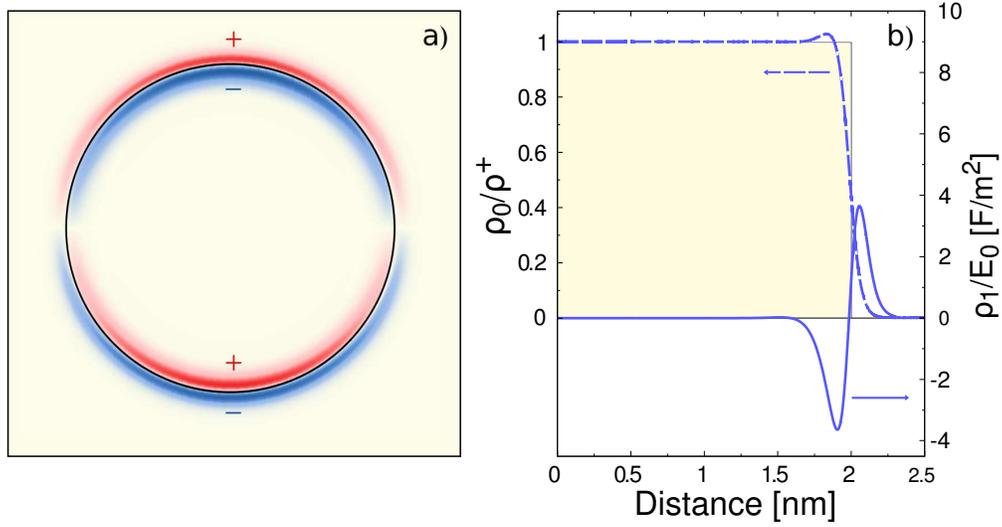}
\caption{Panel a) Charge density distribution $\rho_{\rm 1}(\bf r)$ for the Bennett resonance for a Na cylindrical nanowire of radius
$R=2$\,nm. Same as Figure~\ref{fig:sodium}b), now for the Bennett resonance.}
\label{fig:Bennett}
\end{figure}

The absorption cross section spectrum in Figure~\ref{fig:sodium}a shows
another resonance peak at
$\hbar\omega_{\rm res}= 4.918 \, \rm eV$. This resonance is known in the
literature as ``multipole surface plasmon'' or ``Bennett resonance'', first predicted by
Bennett in 1970~\cite{Bennett:1970}. The charge distribution
$\rho_{\rm 1}$ associated with this resonance is characterized by the fact that
its integral along a direction orthogonal to the surface vanishes
identically~\cite{Chiarello:2000} (see Figure~\ref{fig:Bennett}). The Bennett resonance
is not observable in the LRA and HW-HD models in Figure~\ref{fig:sodium}, since
it is an effect of the non-uniform charge density in the metal and the electron
spill-out in vacuum.\cite{Bennett:1970}.
 We would like to point out that also in this case the spectral position
of the Bennett resonance that we find is in the same range of the results
obtained with other methods~\cite{Bennett:1970,Tsuei:1990,Liebsch:1993,Stella:2013,
Teperik:2013}. 

Figure~\ref{fig:sodium}c shows again an equilibrium electron density
$n_{\rm 0}$ and an induced charge density $\rho_{\rm 1}$, but now for the
HW-HD model. Consistent with the hard-wall assumption, all charges in
Figure~\ref{fig:sodium}c are indeed localized inside the metal. The equilibrium
density $n_{\rm 0}$ is constant and equal to the local-response value for bulk
Na, whereas the induced density $\rho_{\rm 1}$ can be seen to decrease
monotonically away from the surface, where in local response (not shown) all
induced charge would reside in a delta-function distribution on the interface.
Finally, a comparison of Figures~\ref{fig:sodium}b,c nicely illustrates that
the non-monotonous equilibrium and induced charge densities in the metal region
and the spill-out in the air region are a consequence of taking density
gradients into account in the energy functional of our self-consistent
hydrodynamic theory.

\section{Silver Nanowires - Optical Response}
We will now compare the optical response of Ag nanowires with that of Na
nanowires as discussed above. The aim is to show that our self-consistent HD
theory for both types of metals agrees with experiment and with TD-DFT
calculations.

The free-electron gas in Ag consists of electrons in the 5s-band. Analogous to
the case of Na that we discussed above, their spatial equilibrium
distribution $n_{\rm 0}$ can be calculated within the jellium approximation for
the positive-charge background. Such calculations of $n_{\rm 0}$ have already
been done for noble metals, in DFT calculations for
example~\cite{Liebsch:1993,Kulkarni:2013,Pustovit:2006}.

However, unlike for Na, for Ag the interband transitions from the filled
4d-band to the 5s-band near to the Fermi level often cannot be neglected in the
optical response, so that Ag cannot be described as a pure Drude metal at
optical frequencies. Because of this well-known fact, we will take interband
transitions into account in our SC-HD model for Ag.
In principle, a full-electron calculation could be used to account for interband
transitions. However, this is computationally very demanding and beyond the aims
and scope of our hydrodynamic approach. For this reason, we follow the
prescription proposed by~\citeauthor{Liebsch:1993}\cite{Liebsch:1993}, treating
the 4d-core electrons as an effective polarizable medium that contributes to the
optical response by means of an interband permittivity $\varepsilon_{\rm
inter}$. This medium is assumed to extend up to the first plane of nuclei. In the jellium model, the distance $d$ between the edge of the positive background and the first plane of nuclei amounts to half a lattice spacing, and this parameter can
be obtained from experimental data on the specific metallic system. For example, in the case of the Ag
(111), (001), and (110) faces, $d$ is equal to 1.18, 1.02, and 0.72 \AA, respectively~\cite{Liebsch:1993}.

For a good comparison, we again study nanowires of radius $R= 2 \, {\rm nm}$
that are irradiated by a plane wave with in-plane polarization and amplitude
$E_{\rm 0}=1 \, {\rm V/m}$. For Ag the ion density is $n^+=5.8564
\times 10^{28} \, {\rm m^{-3}}$, corresponding to a plasma frequency
$\hbar\omega_{\rm p}=9.01 \, {\rm eV}$.  For the hydrodynamic parameter $\beta$
for Ag we employ $v_{\rm F}=1.39 \times 10^6 \, {\rm m/s}$. The
interband permittivity
$\varepsilon_{\rm inter}$ for Ag is obtained from ref~\citenum{Rakic:1998}.
In the SC-HD model we fix (rather than fit) the distance parameter at $d=1 \, {\rm
\AA}$, which is of the order of the aforementioned distances between the first
plane of nuclei and the edge of the jellium background\cite{Liebsch:1993}. By
contrast, for the HW-HD model we take $d$ to vanish, because the free
electrons do not spill into  free space in this case \cite{Toscano:2012b}.

\begin{figure}[ht]
\centering \includegraphics{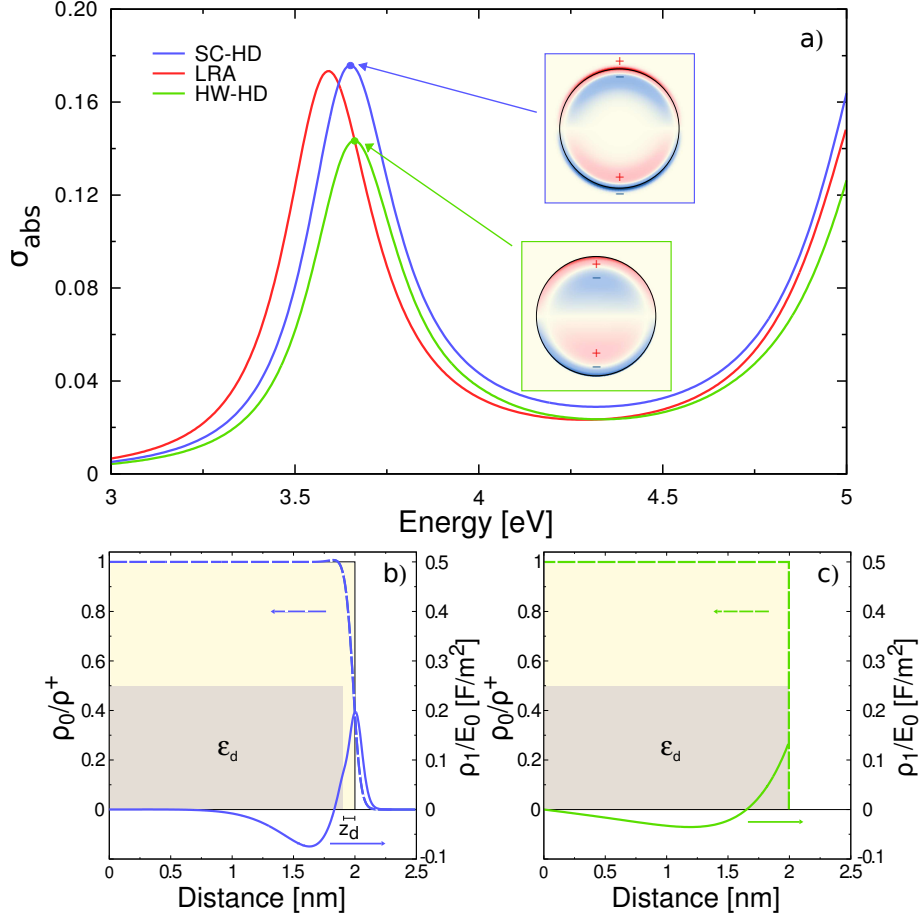}
\caption{Same as Figure~\ref{fig:sodium}, now for Ag nanowires, also with radius $R=2$\,nm.
}
\label{fig:silver}
\end{figure}

Figure~\ref{fig:silver}a shows absorption cross sections for the Ag
nanowire, computed for the same three  models as for the Na nanowire in
Figure~\ref{fig:sodium}. But this time for Ag both the HW- and the
SC-HD models exhibit blueshifts with respect to the
local-response surface-plasmon resonance at $\hbar\omega_{\rm res}=3.594\,{\rm
eV}$. In particular, in the HW-HD model the resonance occurs at
$\hbar\omega_{\rm res}=3.664\,{\rm eV}$ (i.e. blueshifted by $0.0697\,{\rm
eV}$), while for the SC-HD model the resonance is at
$\hbar\omega_{\rm res}=3.653\,{\rm eV}$ (a blueshift of $0.0592 \,{\rm eV}$). As
for Na, the difference in the peak values of the absorption cross section
can be intuitively explained as a fingerprint of the distinct geometrical cross
sections of the scatterers in the three models.

While the blueshift in the HW-HD model is due to the `spill-in' of the
electron density inside the metal, the blueshift in the SC-HD
model
is a more complicated affair, related to the interplay between the 4d-band and
5s-band electrons at the metal-air interface: the induced charge density that
contributes to the SPR oscillates at the unscreened plasma frequency in the
vacuum region ($\omega_{\rm res}=\omega_{\rm p}/\sqrt{2}$), and it oscillates
at
the screened plasma frequency $\omega_{\rm res}=\omega_{\rm
p}/\sqrt{1+\mbox{Re}(\varepsilon_{\rm inter}}$) inside the metal. This complex
mechanism leads to a net increase of the resonance frequency for Ag
nanowires\cite{Liebsch:1993}. Our important point here is that the SC-HD model
is the first semiclassical model to account for this mechanism.

Figure~\ref{fig:silver}b shows the equilibrium electron density $n_{0}$ for
Ag nanowires, exhibiting both the charge spill-out effect and an onset of
Friedel oscillations, quite analogous to the case of Na. The induced
charge density $\rho_{\rm 1}$ is also shown, and although for Ag it is
screened by the polarization effects of the effective medium due to interband
transitions, at least up to a distance $d$ away from the jellium interface, we
again see an oscillation in the internal region, and a peak followed by decay in the free-space region.

Figure~\ref{fig:silver}c shows the constant equilibrium electron density $n_{0}$ and the induced charge
density $\rho_{\rm 1}$ in the HW-HD model. Both are fully localized
inside the nanowire, consistent with the hard-wall assumption. The charge density oscillation is due to
the screening effects of the core electrons, in contrast to the monotonous
spatial decay of $\rho_{\rm 1}$ for Na in Figure~\ref{fig:sodium}c.

Finally, a comparison of Figures~\ref{fig:sodium}a and~\ref{fig:silver}a shows
that the absorption cross section of Ag nanowires increases for higher energies, while
it drops for Na wires. The difference can be attributed to interband
transitions in Ag that have no counterpart in Na. The interband transitions also
prevent the occurrence of a Bennett resonance in Ag nanowires.

\section{Size-dependent frequency shifts}
Until now we studied the SPR shifts for both Na and Ag nanowires with a
fixed radius of $R=2 \, {\rm nm}$, but to support our interpretations we will
now present a systematic study of the dependence of the SPR shifts on
the radius of the nanowire. An advantage of our semiclassical SC-HD model is
that numerical calculations are feasible also for wires of considerably larger
radii. For the HW-HD model in the quasi-static approximation,  analytical
espressions are known for the size-dependent resonance shift, which in that
model is always a blueshift, with a leading $1/R$ scaling.\cite{Raza:2013} We
are now in the position to test whether this scaling behavior is a
consequence of the hard-wall approximation, or also occurs in our SC-HD model
that allows spill-out of the free electrons.

\begin{figure}[ht]
\centering \includegraphics{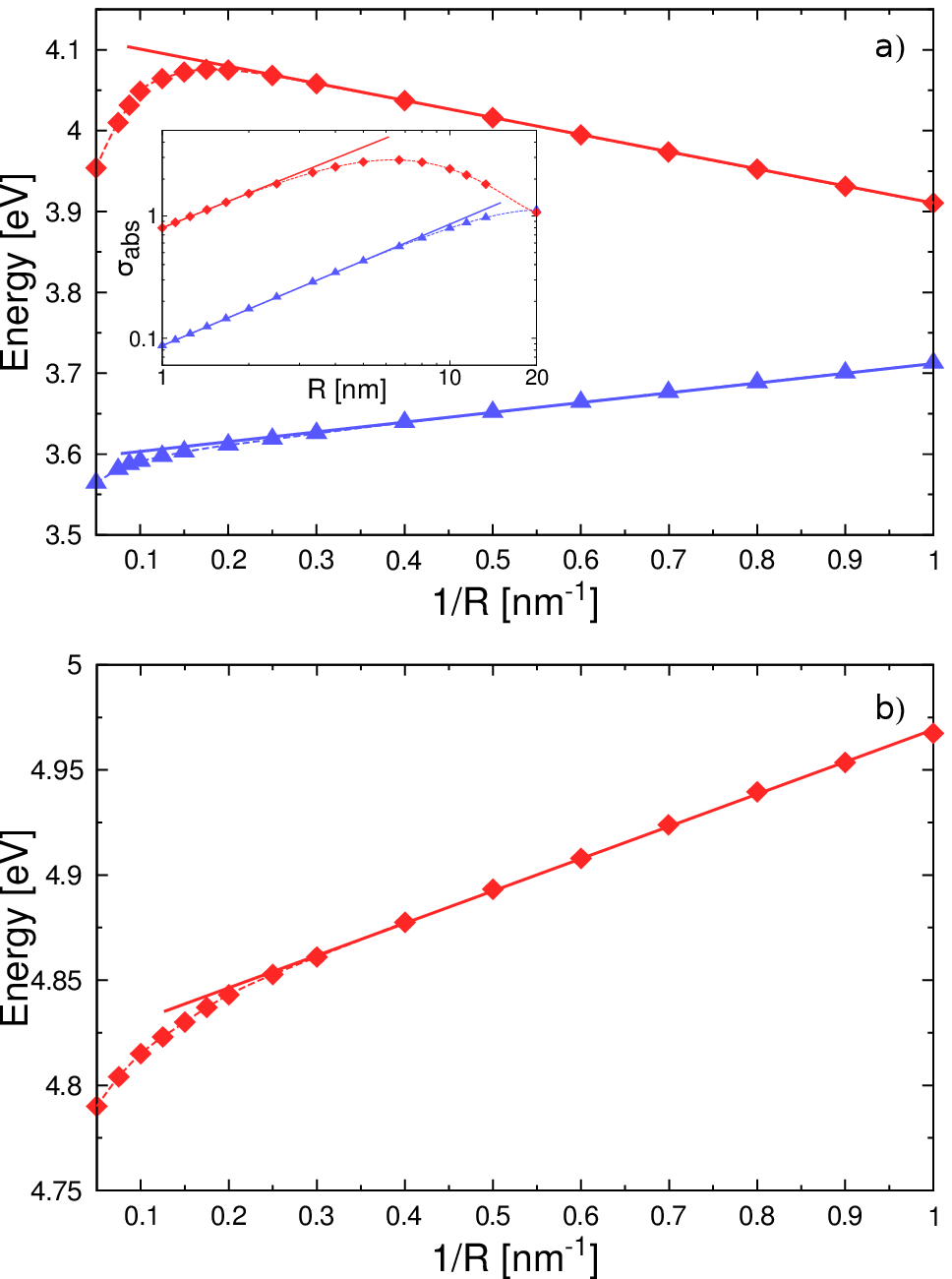}
\caption{Panel a) Surface plasmon resonance frequency versus the inverse radius $1/R$ for a Na nanowire (red squares) and an Ag
nanowire (blue triangles) as calculated in the SC-HD model. The dashed lines are guides to the eye, whereas the full lines illustrate the $1/R$ dependence. Inset: Log-log graphs of the absorption cross
section exactly at the SPR
versus radius $R$ for a Na nanowire (red squares) and an Ag nanowire (blue triangles) in the LRA model. Panel b) ``Bennett" resonance frequency versus the inverse radius $1/R$ for a Na nanowire. }
\label{fig:res_shifts}
\end{figure}

Figure~\ref{fig:res_shifts}a shows the SPR frequency shift as a function
of the inverse radius $1/R$ as computed with the SC-HD model. The red line
represents the data for Na.  For particles of radius $R<5 \, {\rm nm}$, the
resonance frequency varies linearly with $1/R$, with a negative slope of $-0.2
\,{\rm eV}/{\rm nm^{-1}}$, in other words the SPR frequencies
\emph{redshift} when the particle size is decreased. This finding based on our
SC-HD model agrees with the numerical results obtained by Li
\textit{et al.}~\cite{JianHao:2013}, who also predict the $1/R$ scaling by using
TD-~DFT methods.
The blue line in Figure~\ref{fig:res_shifts}a shows the $R$-dependent frequency
shift for Ag. This time the resonance frequency increases linearly with
$1/R$ for wire radii $R<8 \, {\rm nm}$, with slope $0.12 \,{\rm eV}/{\rm
nm^{-1}}$. Thus our SC-HD model predicts that the SPR
frequencies for Ag \emph{blueshift} when decreasing the particle size, in
agreement with the $1/R$ scaling observed experimentally by Charl\'{e}
\textit{et al.}~\cite{Schulze:1989}
So from Figure~\ref{fig:res_shifts}a we can now appreciate that the $1/R$ scaling
is not a consequence of the hard-wall approximation, since it is now also found
in the SC-HD model that includes electron spill-out. Our major result is of course our hydrodynamic (SC-HD)
prediction of the redshift for Na nanowires in Figure~\ref{fig:res_shifts}a,
where the HW-HD model incorrectly predicts a blueshift\cite{Stella:2013,Teperik:2013} (not shown here).

It is important to notice that Figure~\ref{fig:res_shifts}a shows a substantial
deviation from the $1/R$ dependence for Na already starting for  $R>5 \, {\rm
nm}$, which we attribute to retardation. A similar but smaller deviation from
the $1/R$ scaling for Ag can also be observed in
Figure~\ref{fig:res_shifts}, but its onset occurs for slightly larger radii,
namely for $R>8 \, {\rm nm}$. 
To substantiate our claim that the deviations from the $1/R$ scaling in our
SC-HD model are due to retardation, we also present LRA calculations of the
dependence of the SPR peak value for $\sigma_{\rm abs}$  on the particle radius.
In the quasi-static regime this peak value follows a power-law
distribution~\cite{Okamoto:2001}, $\sigma_{\rm abs}\propto R^{\rm k}$, which on
a log-log plot as in the inset of Figure~\ref{fig:res_shifts} would give a
straight line of slope ${\rm k}$.  The clear deviation from such power-law
behavior already for Na wires of radii smaller than 10 nm illustrates that
the effects of the retardation become important for rather thin plasmonic nanowires also
in the LRA. Wire radii of at least tens of nanometers are more typical in nanoplasmonic experiments. 
The main panel in Figure~\ref{fig:res_shifts} thus also nicely
illustrates that the SC-HD model combines two important features:
first, it describes free-electron nonlocal response and spill-out, leading to $1/R$ size-dependent blueshifts for some metals and
redshifts for others; second, it takes retardation fully into account, which
for larger wires redshifts surface-plasmon resonances away from their quasi-static values.

Finally, we consider the Bennett resonance frequency shift as a function
of the inverse radius $1/R$  as computed with the SC-HD model (Figure~\ref{fig:res_shifts}b). For particles of radius $R<4 \, {\rm nm}$, the resonance frequency varies linearly with $1/R$, with a positive slope of $0.14
\,{\rm eV}/{\rm nm^{-1}}$, i.e. the Bennett resonance frequencies
\emph{blueshift} when the particle size is decreased.

\section{Conclusion}
In conclusion, we have shown that the Self-Consistent Hydrodynamic Model is able to describe not only nonlocal response but also electronic spill-out of both noble (Ag) and simple (Na) metals.
Both the signs and the values of the resonance shifts obtained with this method
agree with experimental results and predictions obtained with ab-initio methods. 
The SPR blueshift for Ag versus the redshift for Na are only a consequence of  different well-known parameters: Drude parameters, Fermi velocities, interband permittivities $\varepsilon(\omega)$, and the spacing $d$ between the jellium edge and the first plane of nuclei. 

Moreover, Bennett resonances could be identified here for the first time in a
semiclassical model. We also systematically studied the 
size dependence of the Bennett resonance, which we believe has not been done before by other methods. We found that the Bennett resonance frequencies \emph{blueshift} when the particle size is decreased.

The SC-HDM was obtained by extending state-of-the-art hydrodynamic models
considered in nanoplasmonic research to consistently include gradients of the
density into the energy functional that governs the semiclassical dynamics.
Thus the inability of previous hydrodynamic models to find redshifts of surface
plasmon resonances for nanoparticles consisting of simple metals cannot be held
against hydrodynamic models as such. Rather, it was a consequence of using too
simple energy functionals that neglected terms involving gradients of the
electron density, terms which become important near metal-dielectric interfaces,
being most pronounced for Na while spectrally of minor importance for Ag and Au nanowires.

Moreover, the SC-HDM model includes the effects of retardation, that are not
treated in TD-DFT, unless more complicated schemes are
adopted\cite{Vignale:2004}. It provides a fast, flexible and reliable tool that
gives the exciting possibility to treat electronic spill-out and nonlocal
response in relatively large-size systems of arbitrary shapes that are of
interest in the nanoplasmonic applications. Think for example of large dimers with small gap sizes.

For Ag nanowires we find quite good agreement between plasmonic resonances
in the SC-HD model and in the simpler HW-HD model that neglects
spill-out, justifying the use of the simpler model when not focusing on
spill-out. We maintain that this good agreement for Ag (and the not so good agreement for Na) found here is not fortuitous. Rather, the agreement becomes systematically better for metals with higher work functions. 

Our main goal has been to develop a versatile hydrodynamic theory and numerical method that account for both retardation and electronic spill-out, and here we have shown how this can be done. We do not claim that our energy functional is unique, and we stressed that other self-consistent hydrodynamic theories are possible (see Supplementary Material and refs.~\citenum{Vignale:1997,Vignale_book}). We have not tried to be as accurate as possible by fitting density-functional parameters either to experimentally measured values or to more microscopic calculations. For example, one could add fitting functions to the energy functional, fit them for a simple geometry, and use them for more complex geometries. Indeed there is plenty of room to obtain even better agreement with more advanced theories. But exactly by not doing such fits, the good agreement that we already find here is interesting and promising.

In the future we will employ our SC-HD model to other more complex geometries. While in the present work we have mainly focused on size-dependent frequency shifts, we envisage that in the future this model can be extended to
also include size-dependent damping~\cite{Mortensen:2014}.

\begin{acknowledgement}

The authors thank Dr. Shunping Zhang from Wuhan University, Prof. Shiwu Gao from the University of Gothenburg, Prof. Javier Aizpurua from the Donostia International Physics Center (DIPC), as well as Prof. Kristian S. Thygesen, S{\o}ren
Raza, and Dr. Wei Yan from the Technical University of Denmark  for inspiring discussions.
The Center for Nanostructured Graphene is sponsored by the Danish National Research Foundation, Project DNRF58.
This work was also supported by the Danish Council for Independent Research - Natural Sciences, Project 1323-00087.

\end{acknowledgement}

%\begin{suppinfo}

%This will usually read something like: ``Experimental procedures and
%characterization data for all new compounds. The class will
%automatically add a sentence pointing to the information on-line:

%\end{suppinfo}

\bibliography{HD_spillout}

\end{document}